\documentstyle{mn}
\input psfig.sty
\newcommand{\ltsima} {$\; \buildrel < \over \sim \;$}
\newcommand{\gtsima} {$\; \buildrel > \over \sim \;$}
\newcommand{\lta} {\lower.5ex\hbox{\ltsima}}
\newcommand{\gta} {\lower.5ex\hbox{\gtsima}}

\def\refitem{\par\parskip 0pt\noindent\hangindent 20pt}

\title[Thermalization in compact sources]
{Thermalization by synchrotron absorption in compact sources: electron and 
photon distributions}

\author[G. Ghisellini, F. Haardt, R. Svensson]
{G. Ghisellini$^1$, F. Haardt$^2$ and R. Svensson$^3$ \\
$^1$ Osservatorio Astronomico di Brera, V. Bianchi, 46, I-22055 Merate (LC), 
Italy\\
$^2$ Dipartimento di Fisica, 
Universit\`a degli Studi di Milano, V. Celoria 16, I-20133 Milano, Italy \\
$^3$ Stockholm Observatory, Saltsj\"obaden, Sweden }

\date{Accepted 1997 December 12; received ***; in original form ***}

\begin{document}

\maketitle

\begin{abstract}

The high energy continuum in Seyfert galaxies and galactic black hole
candidates is likely to be produced by a thermal plasma. There are difficulties
in understanding what can keep the plasma thermal, especially during fast
variations of the emitted flux. Particle--particle collisions are too
inefficient in hot and rarefied plasmas, and a faster process is called for. We
show that cyclo--synchrotron absorption can be such a process: mildly
relativistic electrons thermalize in a few synchrotron cooling times by
emitting and absorbing cyclo--synchrotron photons. The resulting equilibrium
function is a Maxwellian at low energies, with a high energy tail when Compton
cooling is important. Assuming that electrons emit completely self absorbed
synchrotron radiation and at the same time Compton scatter their own
cyclo--synchrotron radiation and ambient UV photons, we calculate the time
dependent behavior of the electron distribution function, and the final
radiation spectra. In some cases, the 2--10 keV spectra are found to be
dominated by thermal synchrotron self--Compton process rather than by thermal
Comptonization of UV disk radiation. 

\end{abstract}


\section{Introduction}

Particle--particle collisions among electrons, and electrons and ions, are rare
in rarefied hot plasmas, such as the ones thought to be responsible for the
high energy emission in Seyfert galaxies. It is therefore unclear if the plasma
in these sources can be described by a thermal, Maxwellian distribution,
especially during the fast variations often observed in their X--ray flux and
spectrum. On the other hand the OSSE observations of the brightest Seyfert
galaxies (i.e. NGC 4151 and IC 4329, see e.g. Zdziarski, Johnson \& Magdziarz
1996, Madejski et al. 1995) show the presence of a high energy cut--off, highly
indicative of a thermal nature of the underlying plasma. Similar indications
are  obtained from the sum of 60 combined spectra of 27 Seyfert galaxies
(Zdziarski et al. 1995). 

In this paper, we propose a new thermalization mechanism, based on the process
of synchrotron and cyclotron self--absorption. Electrons embedded in a magnetic
field emit and absorb their own cyclo--synchrotron radiation and can exchange
energy by exchanging photons. We will show that complete thermalization occurs
(in a few cooling times) in sources magnetically dominated and when the
radiation is completely self--absorbed. Increasing the amount of Compton
cooling makes the peak of the Maxwellian distribution to shift to lower
energies, with the development of a high energy tail. 

Using the found equilibrium electron distribution we compute the Comptonization
spectra in the framework of the two--phase model of Haardt \& Maraschi (1991)
and Haardt, Maraschi \& Ghisellini (1994). We calculate how the hot magnetized
plasma, above a relatively cold accretion disk, reaches a steady state
equilibrium distribution through synchrotron self--absorption and by
Comptonizing the seed photons coming from the disk. We show that as long as the
hot phase is magnetically dominated, the equilibrium temperatures and mean
energies of the emitting plasma are in agreement with the existing
observations, quite independently of the size of the region. There is however
an important difference with respect to a pure thermal scenario, in which
particles are assumed to always be in thermal equilibrium. Here, in fact we
consider the case of continuous injection of new, energetic electrons, up to
some maximum energy $\gamma_{\rm max}m_{\rm e}c^2$. This implies that: 

1) To reach steady state requires some form of escape of the injected electrons
in order to avoid a pile up of cold electrons. Also reacceleration could occur,
but we do not consider this case here. 

2) The final distribution is sensitive to the mean energy of the injected
electrons, i.e. to $\gamma_{\rm max}$, and to the spectral shape of the
injected distribution. 

3) We neglect the possible role played by electron--positron pairs created by
the produced $\gamma$--rays. This could be a self--regulating mechanism to keep
$\gamma_{\rm max}$ small (see Ghisellini, Haardt \& Fabian, 1993). Here we
assume a given fixed value of $\gamma_{\rm max}$. 

Note that the model proposed here is an hybrid between non--thermal and thermal
models, because it assumes a continuous injection of electrons (and not a
re--heating), which nevertheless reach a thermal (or quasi--thermal)
distribution. 

Previous related studies on the subject of synchrotron reabsorption have been
made by Ghisellini, Guilbert \& Svensson (1988, hereafter GGS88) by Ghisellini
\& Svensson (1989),  and by de Kool, Begelman \& Sikora (1989). These studies
demonstrated that a quasi--Maxwellian distribution can form at the low energy
end of an otherwise power law distribution. In these cases, in fact, electrons
were injected at very high energies ($\gamma_{\rm max}\ge 10^3$) and thin
synchrotron emission was the main cooling process. The problem of deriving the
equilibrium distribution in the presence of self--Compton losses was briefly
discussed by Ghisellini \& Svensson (1989). 

The present paper differs from previous work because we study a regime where
cyclo--synchrotron emission is almost completely self--absorbed, and the main
cooling mechanism is due to the Inverse Compton of photons produced in a
relatively cold accretion disk. 

The structure of the paper is as follows: in Section 2 we describe the main
assumptions of the model and the used equations, in Section 3 we describe our
results, which are then discussed in Section 4. 

\section{Set up of the system}

Assume that in a region of dimension $R$ relativistic electrons are injected at
a rate $Q(\gamma)$ [cm$^{-3}$ s$^{-1}$] between $\gamma_{\rm min}$ and
$\gamma_{\rm max}$. A tangled magnetic field $B$ of energy density $U_{\rm B}$
makes them radiate synchrotron (S) photons. These photons, together with
photons produced externally to the region, interact with the electrons by the
Inverse Compton (IC) process. If the electron distribution extends to a
$\gamma_{\rm max}$ of the order of a few, the synchrotron spectrum is
completely self absorbed, and the total radiation energy density ($U_{\rm r}$)
may be dominated by the photons produced externally. We will assume $U_{\rm
B}\gg U_{\rm r}$. Due to self absorption and the presence of external photons,
this does $not$ correspond to the prevalence of the synchrotron luminosity over
the inverse Compton one. 

The electron distribution $N(\gamma, t)$ is the result of the injection, the S
and IC losses, and the energy gain due to self absorption (GGS88) 

$$
{\partial N \over \partial t} \, =\,
{\partial \over \partial \gamma}
\left[ (\dot \gamma_{\rm S}+\dot \gamma_{\rm C}) N + H \gamma p
{\partial \over \partial \gamma }
\left( N\over \gamma p \right) \right] + Q - {N \over t_{\rm esc}}\ ,
\eqno(1)
$$
where the momentum $p$ is measured in units of $m_{\rm e}c$ and
$\dot\gamma_{\rm S}$ and $\dot \gamma_{\rm C}$ are the cooling rates for
synchrotron and Inverse Compton losses, respectively: 

$$
\dot \gamma_{\rm S} \, =\, {4\over 3} {\sigma_{\rm T} c p^2 U_{\rm B} 
\over m_{\rm e}c^2}\ ,
\eqno(2)
$$

$$
\dot \gamma_{\rm C}(t) \, =\, {4\over 3} {\sigma_{\rm T} c p^2 U_{\rm r}(t) 
\over m_{\rm e}c^2}\ .
\eqno(3)
$$
Equation (3) assumes that the Compton cooling is in the Thomson regime. The
kinetic equation (1) differs from the analougous equation (2) of GGS88 by the
inclusion of the Compton loss term $\dot \gamma_{\rm C}$. The factor $H$ is
defined below. The last term represents electron escape from the plasma region
with the escape time being $t_{\rm esc} = R/v_{\rm esc}$. We use $v_{\rm
esc}=c$.

We assume that the radiation energy density $U_{\rm r}$ is dominated by an
external soft photon distribution, and by the `hard' radiation produced by
Comptonizing these photons. The soft photons are assumed to arise from the
reprocessing of half of the hard radiation  by cold matter in the vicinity of
the active region  as detailed in Haardt \& Maraschi (1991). In steady state,
all the power injected in the form of relativistic electrons balances the
escaping luminosity. Part of the power escapes as kinetic energy flux and part
is converted into radiation, mainly at high (UV and X--ray) energies, via IC.
The ratio of the two depends upon the ratio of the escape and the IC cooling
timescales. 

Since the cyclo--synchrotron emission is completely self--absorbed, it is the
IC process which is mainly responsible for the radiation losses, even if
$U_{\rm B}\gg U_{\rm r}$. The Compton process becomes efficient after a time
$\approx 2R/c$, which is the timescale required to build up the soft photon
radiation field. After this time the radiation energy density can be written as
$$
U_{\rm r}(t)\, =\, a{R\over c}\,(1+\tau_{\rm T}) 
m_{\rm e}c^2\int_{\gamma_{\rm min}}^{\gamma_{\rm max}} 
\left[ Q(\gamma) - {N(\gamma) \over t_{\rm esc}} \right] (\gamma-1) d\gamma \ ,
\eqno(4)
$$ 
where $a$ is a numerical coefficient (of order unity) which depends on the
geometry ($\sim 3/4$ for a sphere and $\sim 1/2$ for a slab). We use $a=3/4$.
The factor $(1+\tau_{\rm T})$ accounts for the enhancement of the photon
density due to Thomson scattering with the electrons, the optical depth of
which is 
$$
\tau_{\rm T}(t) \, =\, \sigma_{\rm T} R \int_1^{\gamma_{\rm max}} 
N(\gamma, t)d\gamma\ .
\eqno(5a)
$$
At equilibrium, the number of injected electrons must be balanced by the number
of escaping electrons. In steady state the optical depth becomes 
$$
\lim_{t\to\infty} \tau_{\rm T}(t) \, =\, \sigma_{\rm T}Rt_{\rm esc} 
\int_{\gamma_{\rm min}}^{\gamma_{\rm max}} Q(\gamma)d\gamma\ .
\eqno(5b)
$$

The factor $H(\gamma, t)$ in equation (1) describes the heating of the
electrons  and their diffusion in energy. It is defined as 
$$
H(\gamma, t) \, =\, {1 \over 2m_{\rm e}^2c^2} \int_0^\infty 
{J(\nu, t) \over \nu^2}
j_{\rm s}(\nu ,\gamma) d\nu\ ,
\eqno(6)
$$
where $j_{\rm s}(\nu ,\gamma)$ is the power spectrum emitted by a single
electron through the synchrotron process, and $J(\nu, t)$ is the radiation mean
intensity due to the S and IC process, with the former dominating the
absorption (and hence the heating). $J(\nu, t)$ is calculated by 
$$
J(\nu ,t)\, =\, 
{\mu (\nu ,t)\over \kappa ( \nu ,t )}\,
\left[1-e^{-R\kappa (\nu ,t)}\right]
\eqno(7)
$$
where $\mu(\nu , t)$ and $\kappa (\nu, t)$ are the emissivity and the
absorption coefficient of cyclo-synchrotron radiation, as calculated in GGS88. 

Since we are dealing with mildly relativistic electrons, we must specify the
cyclo--synchrotron emissivity of the single electron. Most existing expressions
are either nonrelativistic or only valid at large harmonics (Petrosian 1981). A
simple one parameter phenomenological expression for the cyclo--synchrotron
power spectrum is 
$$
j_{\rm s} (\nu ,\gamma )\, =\,
{4p^2\over 3}\,
{\sigma_{\rm T}c U_{\rm B}\over \nu_{\rm B}}\, {2\over 1+3p^2}
\exp\biggl[{2(1-\nu/\nu_{\rm B})\over 1+3p^2}\biggr]\, ,
\eqno(8)
$$
where $\nu_{\rm B}$ is the Larmor frequency. This expression (i) has the
correct cooling rate, equation (2), when integrated over frequency, (ii) has
the correct frequency dependence, $\propto \exp(-2\nu/\nu_B)$, at large
harmonics ($\nu\gg\nu_B$) in the nonrelativistic limit, (iii) has the correct
frequency dependence, at large harmonics in the ultrarelativistic limit, and
(iv) the agreement at smaller harmonics, $\nu/\nu_B<100$, is better than 40\%
at $\gamma =2$. 

For $\gamma> 2$ we use the usual synchrotron formula averaged over
isotropically distributed pitch angles (see, e.g., GGS88). 

The kinetic equation (1) can be solved for $N(\gamma, t)$ as a function of
$Q(\gamma)$, the magnetic field $B$, and the size $R$. Equivalently, besides
$R$, one can specify the injected and the magnetic compactnesses. The total
(including rest mass) injected compactness is 
$$
\ell_{\rm inj}\, =\, {\sigma_{\rm T} \over m_{\rm e}c^3} {L_{\rm inj}  
\over R}\ ,
\eqno(9)
$$
where 
$$
L_{\rm inj}\, =\, V m_{\rm e}c^2 
\int_{\gamma_{\rm min}}^{\gamma_{\rm max}} Q(\gamma)\gamma d\gamma \ ,
\eqno(10)
$$
and where $V$ is the volume of the region. The magnetic compactness is defined
as 
$$
\ell_{\rm B}\, =\, {\sigma_{\rm T} \over m_{\rm e}c^2} R U_{\rm B} \ .
\eqno(11)
$$

In the two phase model for Seyfert galaxies, the high energy radiation could be
produced by a uniform corona (Haardt \& Maraschi 1991) or by localized active
regions (possibly highly magnetized; Haardt, Maraschi \& Ghisellini, 1994).
Assuming the second possibility, the dimensions of each region, or blob, is of
the order of a few Schwarzschild radii. (i.e. $R\sim 10^{13}$ cm). 

Observations of the high energy continuum and variability in Seyfert galaxies
indicate that their compactness is of the order of 1--300 (Done \& Fabian,
1989). We then assume that $\ell_{\rm inj}$ is in these range of values. Our
assumption of $U_{\rm B}\gg U_{\rm r}$ then also  fixes the range of $\ell_{\rm
B}$. When escape is not important (i.e., for $\ell_{\rm B}\gta 1$, see below)
we have 
$$
{U_{\rm r} \over U_{\rm B}} \, \approx \, {9\over 16\pi} \, 
{\ell_{\rm inj} \over \ell_{\rm B}} \left( 1 - {1\over <\gamma>}\right)
(1+\tau_{\rm T})\ ,
\eqno(12)
$$
where $<\gamma>$ is the average of $\gamma$ over $Q(\gamma)$.

The ratio of the cooling time scale, $t_{\rm cool}= (\gamma-1)/ (\dot
\gamma_{\rm S} + \dot \gamma_{\rm C})$, to the escape time, $t_{\rm esc}$, is
given by 
$$
{t_{\rm cool} \over t_{\rm esc}} \, \approx \, {3 \over 4}
{v_{\rm esc} \over c} {1 \over (1+\gamma) \ell_{\rm B} 
(1+ U_{\rm r}/U_{\rm B})}\ ,
\eqno(13)
$$
This is always less than unity for $\ell_{\rm B} \gta 1 $, implying that
electrons cool rather than escape for sufficiently large compactnesses. Below
we use $\ell_{\rm B} =$ 10 or 30. 

Electrons may also thermalize through Coulomb energy exchange (e.g., Stepney
1983, Dermer \& Liang 1989, Nayakshin \& Melia 1996, Pilla \& Shaham 1997). The
thermal electron-electron Coulomb energy  exchange rate can be approximated
with (e.g., Stepney 1983) 
$$
\dot \gamma_{\rm Coul}(\Theta)\,  \approx \, {3 \over 8} {\tau_{\rm T} c 
\ln \Lambda
\over R \Theta^{1/2} ( \pi^{1/2}+ \Theta^{1/2}) }\ ,
\eqno(14)
$$
where $\ln \Lambda$ is the Coulomb logarithm, typically $\approx$ 10-20, and
$\Theta \equiv kT / m_{\rm e} c^2$ is the dimensionless temperature. The
thermal average of equation (2) becomes 
$$
\dot \gamma_{\rm S}(\Theta) \, \approx \, 
(c/R) (4/3) \Theta (1 + 4 \Theta) \ell_{\rm B}\ .
\eqno(15)
$$
Thermalization by synchrotron selfabsorption dominates when $\dot \gamma_{\rm
S} > \dot \gamma_{\rm Coul}$, which gives $\ell_{\rm B} >$ $(9/32\pi^{1/2})
\tau_{\rm T} \ln \Lambda / \Theta^{3/2}$ for $\Theta < 1$. One sees that the
Coulomb process dominates for small temperatures and large $\tau_{\rm T}$
(i.e., large electron densities).  From equations (5b), (9), and (10), we find
that the Thomson optical depth is given by 
$$
\tau_{\rm T} = {3 \over 4 \pi} {v_{\rm esc} \over c} 
{\ell_{\rm inj} \over <\gamma>}\ .
\eqno(16)
$$
Thermalization by synchrotron selfabsorption then dominates for temperatures 
$$
\Theta >  0.11 \left({\ln \Lambda \over <\gamma>} 
{\ell_{\rm inj} \over \ell_{\rm B}}\right)^{2/3}\ .
\eqno(17)
$$

\section{Results}
\subsection{Time evolution}

Figure 1 shows the evolution of the electron distribution towards the
equilibrium configuration. The injected electrons have a Gaussian energy
distribution peaking at $\gamma=10$. The magnetic field has a value $B\simeq
5.5\times 10^3$ G and the injected power corresponds to $\ell_{\rm inj}=1$. The
size of the source is $R=10^{13}$ cm. As can be seen, the equilibrium shape of
$\tau(p)\equiv \sigma_{\rm T} R N(p)$ is reached in a time $\sim 0.1 R/c$,
about equal to the nonrelativistic cooling time (see eq. 13). With the assumed
input parameters, the synchrotron terms (emission,  absorption and energy
diffusion) in the kinetic equation are dominant over  Compton losses. Gains and
losses in this case almost perfectly balance.  As a result the equilibrium
electron distribution is a Maxwellian. Figure 1  also shows that the high
energy part of the Maxwellian distribution is formed  earlier than the low
energy part, due to the higher efficiency of  exchanging photons of the high
energy electrons. A slower evolution takes place after 0.1(R/c), as balance
between electron injection and electron escape is achieved on a time scale of a
few $t_{\rm esc}$. Only then have both the shape and the amplitude of $N(p)$
reached their equilibrium values.

\begin{figure}
\psfig{file=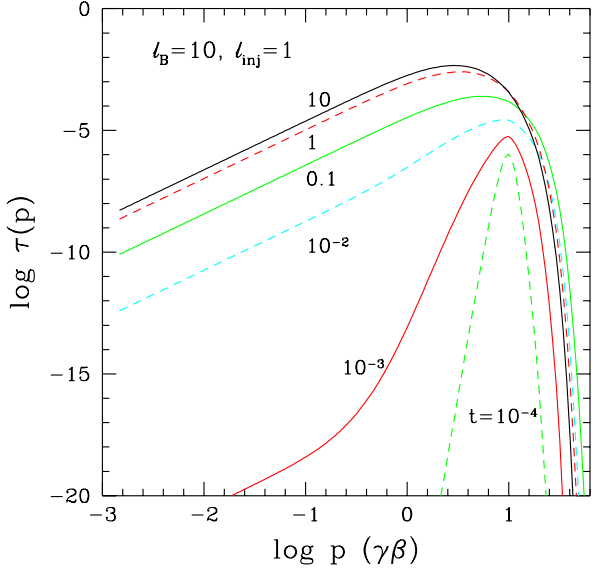,width=8.5truecm,height=8.5truecm}
\caption[h]{Evolving electron distribution at different times 
(measured in units of $R/c$), as labelled.
The size of the source is $R=10^{13}$ cm, $\ell_{\rm B}=10$, and
$\ell_{\rm inj}=1$. The injected distribution is a Gaussian centered at 
$\gamma=10$.}
\end{figure}

\subsection{Steady equilibrium distributions}

The equilibrium distributions for different values of the injected power are
shown in Figure 2. The magnetic compactness is set to $\ell_{\rm B}=30$,
corresponding to $B=9.6\times 10^3$ G for $R=10^{13}$ cm. In all cases, the
injected distribution is $Q(\gamma)=Q_0\, p/\gamma \, \exp(-\gamma/\gamma_{\rm
c})$, where $Q_0$ is a normalization constant and $\gamma_{\rm c}=3.33$. The
minimum Lorentz factor is $\gamma_{\rm min}=1$. The resulting mean injected
Lorentz factor is $<\gamma>\simeq 4.6$. As long as $\ell_{\rm inj} \ll
\ell_{\rm B}$, the distribution is a quasi--perfect Maxwellian at all energies.
This is the result of the quasi--perfect balance between energy gains and
losses, as also seen in Fig. 2: Compton losses (not balanced by a corresponding
Compton heating term) are a small perturbation. As $\ell_{\rm inj}$ increases
towards values $\simeq \ell_{\rm B}$, Compton losses start to be relevant,
competing with synchrotron processes. At high energies, losses overcome gains,
and the electrons diffuse backwards in energy, until the subrelativistic regime
is reached. Only in this regime, the increased efficiency of synchrotron gains
(relative to losses) halts the systematic backward diffusion in energy, and a
Maxwellian can form (see Ghisellini \& Svensson 1989). We can evaluate the
temperature of this part of $N(\gamma)$ [or, equivalently, of $N(p)$] by
fitting a Maxwellian to the the low energy part of the distribution, up to
energies just above the peak. 

\begin{figure}
\psfig{file=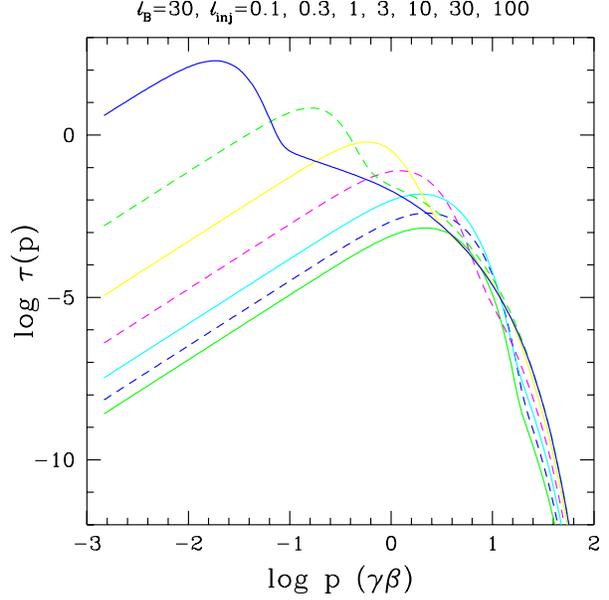,width=8.5truecm,height=8.5truecm}
\caption[h]{Equilibrium electron distributions 
for different injected compactnesses (decreasing from top to
bottom). $R=10^{13}$ cm and $\ell_{\rm B}=30$ are assumed.}
\end{figure}

The resulting temperatures are plotted in Figure 3 as a function of $\ell_{\rm
inj}$. As can be seen, for low values of $\ell_{\rm inj}$ the temperature is
quasi--constant, and decreases for $\ell_{\rm inj}$ greater than 1. Also shown
in Figure 3 is the behavior of the ``effective temperature" $T_{\rm eff}$,
defined by [$\Theta_{\rm eff}\equiv kT_{\rm eff}/(m_{\rm e}c^2)$]: 
$$
16\Theta_{\rm eff}^2 + 4\Theta_{\rm eff} \, =\, {4\over 3}<\gamma^2-1>_{_N}\ ,
\eqno(18)
$$
where the average is over the electron distibution $N(\gamma)$. As expected,
for small $\ell_{\rm inj}$, $T_{\rm eff}$ is well approximated by $T$, while
the effective temperature is larger than $T$ for $\ell_{\rm inj}\gta 10$, as a
result of the shift of the Maxwellian part to small energies, and the
corresponding appearence of the high energy tail. 

\begin{figure}
\psfig{file=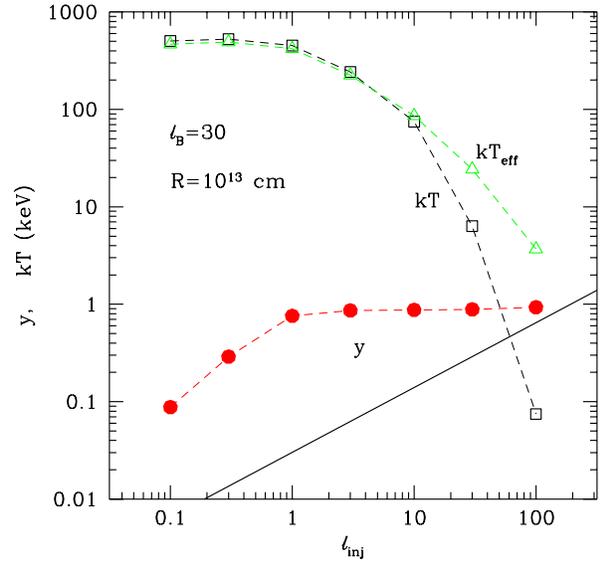,width=8.5truecm,height=8.5truecm}
\caption[h]{Temperatures and effective temperatures of the distributions
shown in Fig. 2. Also shown are the corresponding 
values of the Comptonization $y$ parameter. Above the solid
line, synchrotron self absorption dominates over Coulomb
exchange as the thermalization mechanism. }
\end{figure}

Along the sequence, the optical depth increases from $\tau_{\rm T}=5\times
10^{-3}$ for $\ell_{\rm inj}=0.1$ to $\tau_{\rm T}=5$ for $\ell_{\rm inj}=100$,
according to equation (16). For our parameters and $\ln \Lambda = 20$, equation
(17) becomes $\Theta > 0.03 (\ell_{\rm inj})^{2/3}$, which is plotted as the
solid line in Figure 3. One sees that  that synchrotron self absorption
dominates the thermalization for all cases with $\ell_{\rm inj}$ smaller than
about 60 . For the case $\ell_{\rm inj} =$ 100, one cannot neglect Coulomb
thermalization. 

With the values of $\Theta_{\rm eff}$ and $\tau_{\rm T}$, it is of interest to
calculate the Compton parameter $y$ defined by 
$$
y \, =\, (\tau_{\rm T}+\tau_{\rm T}^2)(16\Theta^2_{\rm eff}
+4\Theta_{\rm eff})\ ,
\eqno(19)
$$
which is also plotted in Figure 3. At equilibrium, $\tau_{\rm T}$ is given by
equation (16) and $\Theta_{\rm eff}$ cannot exceed the value of $\Theta_{\rm
eff}$ obtained by doing the average in equation (18) over  the injected
distribution $Q(\gamma)$. This implies that $y$ has a maximum possible value,
$y_{\rm max}$. This limiting value is reached only when radiative cooling is
negligible. The actual value of $y$ is then further constrained by the amount
of radiative cooling. 

For $\ell_{\rm inj}\ge 1$, $y_{\rm max}\gg 1$, but the amount of Compton
cooling (dominant in this cases) keeps the value of $y$ close to unity. The
fact that $y \sim 1$ is to be expected: this value ensures equal power between
the Comptonized radiation and the soft seed photon emission. 

For $\ell_{\rm inj}\le 1$, $y_{\rm max}$ is instead less than unity: Compton
cooling is negligible, and the actual value of $y$ is only slightly less than
$y_{\rm max}$, due to synchrotron losses, which, albeit small, are in these
cases dominant over Compton losses.

\subsection{Spectra}

In Figure 4, we show the radiation spectra corresponding to some of the 
equilibrium electron distributions of Figure 2. Each spectrum consists of
several continuum components: 

{i)} the self--absorbed synchrotron spectrum (S);\par

{ii)} the Comptonized synchrotron spectrum (SSC);\par

{iii)} the thermal soft component (bump);\par

{iv)} the spectrum resulting from Comptonization of the photons from
the thermal bump (IC);\par

{v)} the Compton reflection component.

\begin{figure*}
\psfig{file=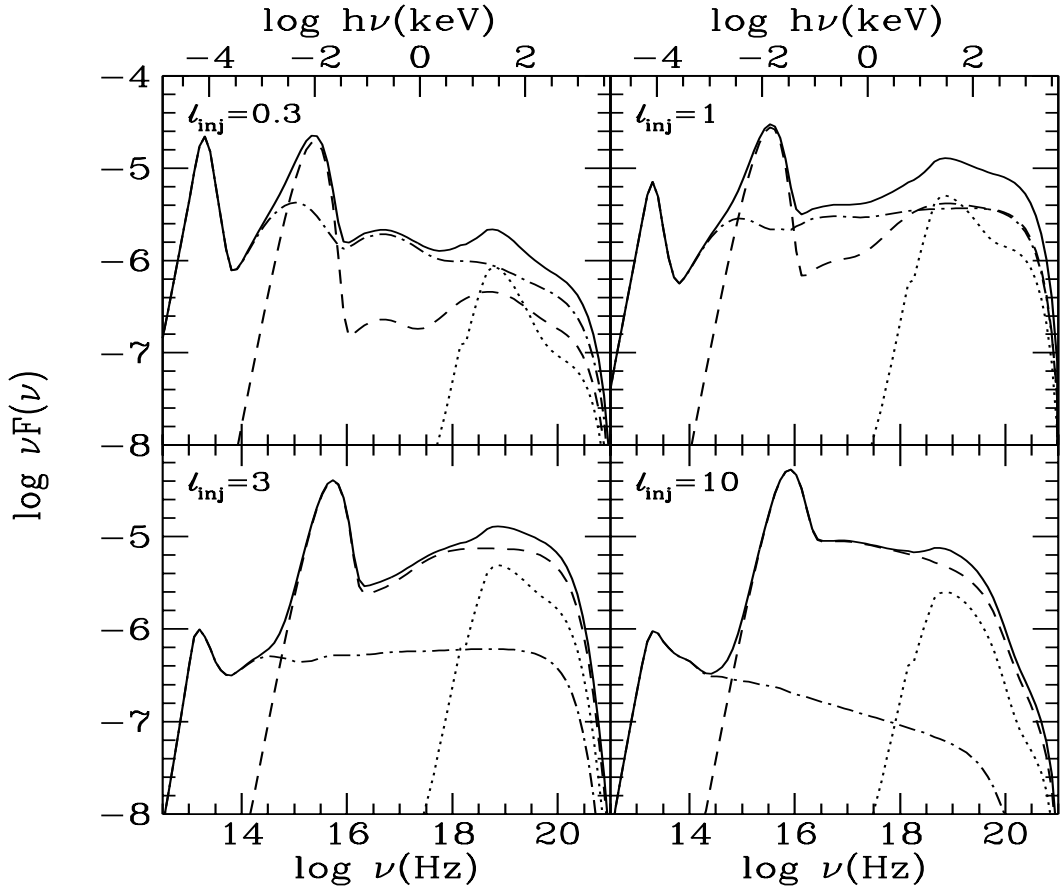,width=17truecm,height=14truecm}
\caption[h]{Radiation spectra calculated with the electron distributions
shown in Fig. 2. 
Dashed line: reprocessed thermal bump and IC components. 
Dash--dotted line: synchrotron and SSC components. Dotted line: 
Compton reflection component. Solid line: total spectrum.}
\end{figure*}

In addition, not shown in the plots, fluorescence line emission from
photoionized iron in the accretion disk and absorption features due to
partially ionized warm gas along the line of sight are expected to be present
in the spectrum. 

For each model the values of the compactnesses of the Compton (SSC+IC)
component and the synchrotron component are calculated as discussed in \S 2. In
the adopted geometry, neglecting the disk albedo, the compactness corresponding
to the soft photon input is $\ell_{\rm soft}=\ell_{\rm ava}/(1+{\rm
e}^{-\tau_{\rm T}})$, where $\ell_{\rm ava}$ is the compactness available for
radiation production, defined as the difference between the injected and the
escaping $kinetic$ compactnesses, i.e. subtracting the rest mass energy of the
electrons from the total energy. 

The temperature of the soft radiation is then calculated assuming black body
emission from a region of radius $R$. The synchrotron and soft component are
then Comptonized in a plane parallel slab following the prescriptions described
in Haardt (1994). For the synchrotron component we assume a homogeneous source
function within the scattering medium, while the thermal soft photon input is
localized below the active region, giving rise to anisotropic Compton emission
(Haardt 1993). A face-on line of sight is assumed for all the spectra. 

The following features may be noticed: 

For $\ell_{\rm inj}\lta 1$, the Comptonized spectrum is bumpy, due to the small
value of $\tau_{\rm T}$ and the large value of $\Theta$. As reported in Figure
3, the $y$ parameter is smaller than unity, making Compton losses relatively
unimportant. The synchrotron component, albeit self--absorbed, is more
important than the Comptonized power. The luminosity in the thermal ``bump" (in
the UV) is roughly half of the synchrotron luminosity (in the IR). Therefore,
SSC and the IC components have approximately the same luminosity ratio as
between the synchrotron and the thermal ``bump" luminosities. \footnote{Note
that in Fig. 4 the thermal ``bump" appears instead as luminous as the
synchrotron component: this is due to the assumed isotropy of the synchrotron
emission, while the ``bump" emissivity has a $\cos i$ angular pattern, where
$i$ is the viewing angle (here, $i=0^{\circ}$). Note also that, while the SSC
emission is isotropic at all energies, the IC component is not: the emission of
the first order scattering in the $i=0^\circ$ direction is depressed (Haardt
1993). At higher energies, both the IC and the SSC components are isotropic,
and here their luminosity ratio has the expected value.} Due to the anisotropy
Compton effect, {\it the 2--10 keV band is dominated by the SSC component},
rather than by the IC. This is in contrast to the general view interpreting the
high energy emission in Seyfert galaxies as due to Comptonization of thermal
bump photons. Here the thermal bump and the X--ray flux are not directly
related. 

Increasing $\ell_{\rm inj}$ up to 1, the Compton cooling becomes more important
($y$ approaches 1), making i) the the synchrotron component (and the related
SSC) weaker, and ii) the hard X--ray slope flatter. 

For $\ell_{\rm inj}\gta 3$ the Comptonized spectra have the typical IC power
law shape, with increasing (steeping) spectral indices as $\ell_{\rm inj}$
increases. The reason for this steepening even for constant $y=1$, is explained
in detail in Haardt, Maraschi \& Ghisellini (1997). 

Finally, note that for $\ell_{\rm inj}\lta 3$, the high energy part can be
described by an exponential, since the electron distribution is a
quasi--perfect Maxwellian in the entire energy range. For $\ell_{\rm inj}\gta
3$, the electron distribution is more complex (see Fig. 2), resulting in a more
complex radiation spectrum above $\sim$500 keV. 

\subsection{Stellar compact sources}

\noindent
The relevant quantities for the synchrotron thermalization process are the
injected and the magnetic compactnesses. These parameters are independent of
radius, and therefore the process should be largely independent upon the size
of the source. However, since $B\propto (\ell_{\rm B}/R)^{1/2}$ (equation 11),
the same value of $\ell_{\rm B}$ now implies $B=9.6\times 10^6$ Gauss. In turn,
the increase of the magnetic field implies that the energy range of the
electrons emitting optically thick synchrotron radiation is reduced. This
affects the synchrotron emission for small values of $\ell_{\rm inj}$, which
now can have an optically thin component from the most energetic electrons. 
\begin{figure*}
\psfig{file=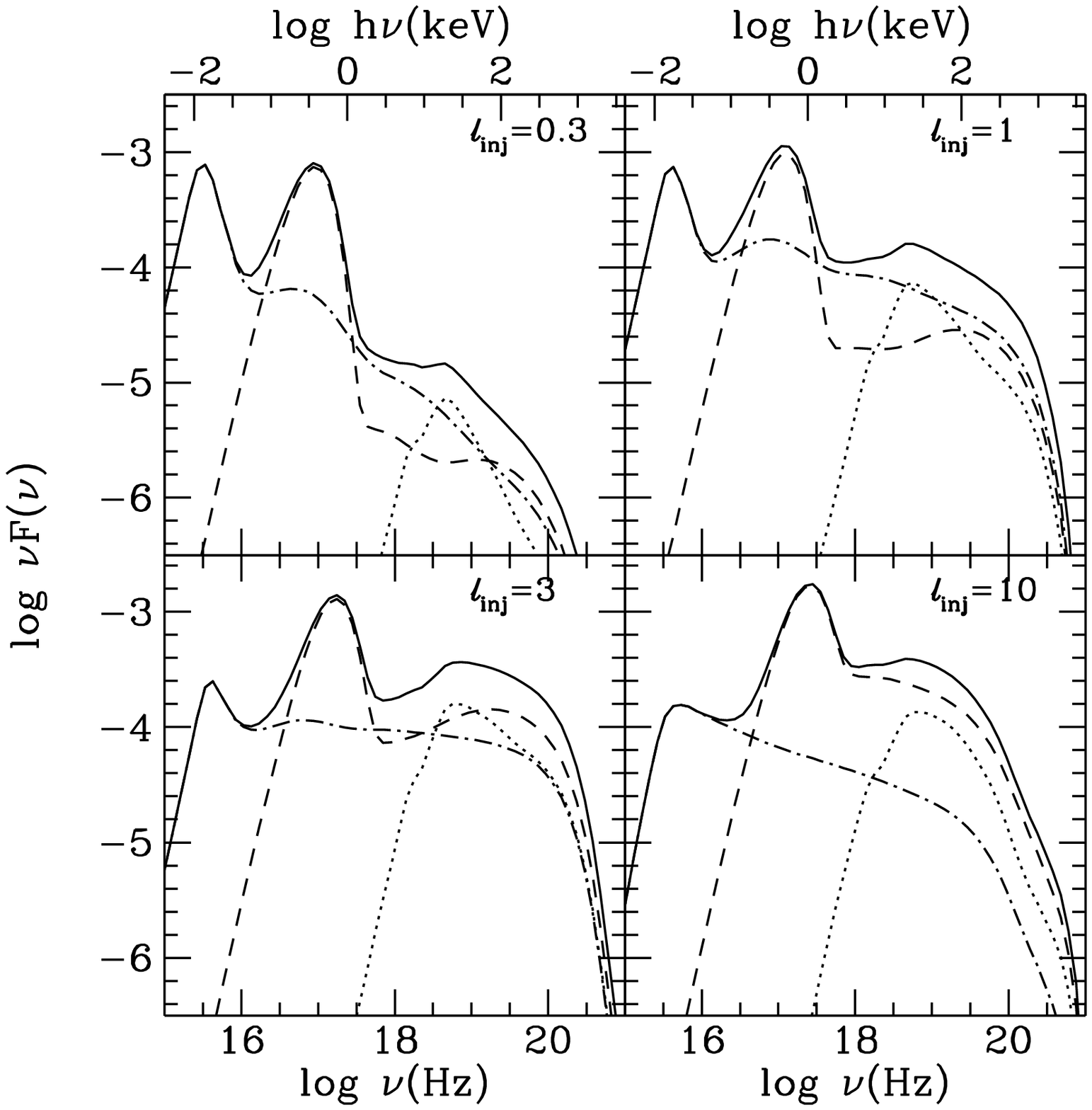,width=17truecm,height=14truecm}
\caption[h]{Radiation spectra for the same values of $\ell_{\rm B}$ and 
$\ell_{\rm inj}$ as in Fig. 4, but for $R=10^7$ cm. 
Dashed line: reprocessed thermal bump and IC components. 
Dash--dotted line: synchrotron and SSC components. Dotted line: 
Compton reflection component. Solid line: total spectrum.}
\end{figure*}
For illustration, we have repeated our calculations assuming a source size of
$10^7$ cm, i.e. few Schwarzchild radii for a galactic black hole. For
$\ell_{\rm inj}\gta 1$ the resulting equilibrium electron distributions are
almost identical to the $R=10^{13}$ case, since the synchrotron radiation is
completely self--absorbed. For $\ell_{\rm inj}\lta 1$, the results are similar,
but the resulting effective temperatures are smaller (by a factor 2 for the
$\ell_{\rm inj}=0.1$ case), due to the presence of a thin synchrotron
component. 

The computed radiation spectra are shown in Figure 5. They exhibit the same
basic features as seen in the extragalactic case (Fig. 4), but with two major
differences: i) as expected, the synchrotron emission is more important, with a
corresponding more relevant SSC component contributing to the X--ray spectrum
for $\ell_{\rm inj}\lta 3$. The increased synchrotron cooling makes the
Comptonized components steeper than the corresponding cases in Figure 4; ii)
the peak energies of the thermal ``bump" and the synchrotron emission are now
in the soft X--ray and UV bands, respectively. 
 
\section{Discussion and conclusions}

We have shown that the exchange of cyclo--synchrotron photons in a self
absorbed source is a very efficient thermalizion mechanism, especially when the
density and temperature of the plasma prevent the thermalization via
particle--particle processes. 

In our calculations we assume that the magnetic energy density in the flaring
corona dominates over both gas and radiation pressure. This is sensible
assumptions in the picture of an active corona. In fact the accretion mechanism
is thought to amplify the magnetic field via some sort of dynamo process, and
eventually particles in the corona are energized via buoyancy and reconnection
of the field lines. Regardless of the largely unknown physics of such
processes, it is conceivable that the particles, (and the radiation field they
create because of cooling), can not be energetically dominant, as their energy
is ultimately drained from the (amplifield) magnetic field. Note that this
argument is true only if the reconnection timescale is larger than the cooling
timescale for a radiating electron, something we have implicitly assumed
allowing for steady state particle injection. 

We now discuss how the main results of the model change by varying some of the
input parameters. 

As already mentioned, we have assumed, for all shown cases, an injected
electron distribution $Q(\gamma)\propto \times p/\gamma\,
\exp(-\gamma/\gamma_{\rm c})$ with $\gamma_{\rm c}=3.33$, resulting in
$<\gamma>\simeq 4.6$. The emitted radiation spectrum is basically controlled by
$\tau_{\rm T}$, since its value, coupled to the condition $y\simeq 1$,
determines the physical conditions in the source. Since $\tau_{\rm T} \propto
\ell_{\rm inj}/<\gamma>$ (eq. 16), the assumed {\it shape} of the  injected
distribution is not important, as long as different distributions  have the
same $<\gamma>$. Changing $\ell_{\rm inj}$ and $<\gamma>$ by the same factor
hence gives rise to the same photon spectrum, provided that the condition
$\ell_{\rm B}\gg \ell_{\rm inj}$ is matched, and that the synchrotron emission
is mostly self--absorbed. In other words, the same spectra shown in, e.g.,
Figure 4, would be obtained for values of $\ell_{\rm inj}$ different than those
indicated if an input distribution with a different $<\gamma>$ were used. It is
important to note that the value of $<\gamma>$ cannot be taken too large, as it
changes the fraction of self--absorbed synchrotron radiation. As long as the
synchrotron emission is mostly self--absorbed, the spectra obtained for the
same value of $\tau_{\rm T}$ and $\ell_{\rm inj}/<\gamma>$ are indeed
identical. Going to very large $<\gamma>$, the conditions of self-absorption
breaks down: the electrons cool down mainly because of synchrotron losses, 
with neglibible Comptonization, changing completely the resulting radiation
spectrum. 

As can be inferred from the shown cases, the value of $R$ is largely
unimportant: its value determines the peak frequencies of the synchrotron and
bump components, but not the main characteristics of the electron and radiation
spectra. 

The thermalization process described in this paper is operating whenever
energetic electrons and some magnetic field is present, but it becomes of great
interest in the study of the high energy spectrum of compact sources when the
mean energy of the electrons is less than a few MeV and when the magnetic field
is energetically dominant. These conditions ensure that the synchrotron
spectrum is completely self--absorbed, and the entire electron distribution is
influenced by self--absorption. This also ensures that the synchrotron emission
does not overproduce the observed IR (optical) emission observed in
radio--quiet AGN (galactic black hole candidates). However, the
cyclo--synchrotron process does contribute to the IR (optical) emission,
thought to be mainly due to reprocessing of the primary disk radiation by,
e.g., dust. 

Since the same electrons contribute also to the X--ray band, simultaneous
variability in the IR and X--ray bands should clearly identify this component,
allowing the determination of the fraction of synchrotron radiation. Its
relative importance decreases by increasing the injected compactness (see Fig.
4). 

An important outcome of our calculations concerns the X--ray emission. If the
magnetic field is dominant (i.e. its energy density is larger than the
radiation energy density) and the Comptonization parameter $y$ is smaller than
unity, the scattering of internal cyclo--synchrotron photons is more important
than the scattering of the external, ``disk" photons. The variability pattern
expected in our model is therefore rather complex, since different components
can contribute in different bands, and their relative importance depends on the
magnetic field dominance. When synchrotron emission is important, the 2--10 keV
emission could be only indirectly related to the UV bump photons. Simultaneous
UV and X--ray variability (though with different amplitudes) should allow the
determination of the importance of the synchrotron and SSC components.

\section*{References}

\refitem Bekefi G. 1966, {\it Radiation Processes in Plasmas} (New York: Wiley)

\refitem de Kool M., Begelman M. C., Sikora M., 1989, ApJ, 337, 66 

\refitem Dermer C. D., Liang E. P., 1989, ApJ, 339, 512

\refitem Done C., Fabian A. C., 1989, MNRAS, 240, 81

\refitem Ghisellini G., Guilbert P., Svensson R., 1988, ApJ, 334, L5 (GGS88)

\refitem Ghisellini G., Svensson R., 1989, Physical processes in hot 
cosmic plasmas, eds. W. Brinkmann, A. C. Fabian \& F. Giovannelli, 
NATO ASI Series, (Kluwer Academic Publishers), p. 395 
 
\refitem Ghisellini G., Haardt F., Fabian A. C., 1993, MNRAS, 263, L9

\refitem Haardt F., 1993, ApJ, 413, 680

\refitem Haardt F., 1994, PhD thesis, SISSA, Trieste

\refitem Haardt F., Maraschi L., 1991, ApJ, 380, L51

\refitem Haardt F., Maraschi L., Ghisellini G., 1994, ApJ, 432, L95

\refitem Haardt F., Maraschi L., Ghisellini G., 1997, ApJ, 476, 620

\refitem Madejski G. M. et al., 1995, ApJ, 438, 672

\refitem Nayakshin S., Melia F., 1996, preprint

\refitem Petrosian V., 1981, ApJ, 251, 727.

\refitem Pilla R. P., Shaham J., 1997, ApJ, in press

\refitem Stepney S., 1983, MNRAS, 202, 467

\refitem Zdziarski A. A., Johnson W. N., Magdziarz P., 1996, MNRAS, 283, 193

\refitem Zdziarski A. A., Johnson W. N., Done C., Smith D., McNaron--Brown K., 
1995, ApJ, 438, L63

\end{document}